\documentclass{optica-article}

\journal{opticajournal} 

\articletype{Research Article}

\usepackage[utf8]{inputenc}
\usepackage{bm}
\usepackage{booktabs}

\newcommand{\vp}{\varphi}
\newcommand{\beps}{\boldsymbol{\varepsilon}}
\newcommand{\Imag}{\operatorname{Im}}
\newcommand{\Real}{\operatorname{Re}}
\newcommand{\diag}{\operatorname{diag}}
\newcommand{\nhat}{\hat{n}}
\newcommand{\dhat}{\hat{d}}

\begin{document}

\title{A unified dielectric-tensor model of macular form birefringence and dichroism}

\author{G.\,P.\ Misson,\authormark{1,*} D.\ Sarenac,\authormark{2} D.\,A.\ Pushin,\authormark{3} and S.\,E.\ Temple\authormark{1,4,5}}

\address{\authormark{1}School of Optometry, Aston University, Birmingham B4 7ET, UK\\
\authormark{2}Department of Physics, University at Buffalo, State University of New York, Buffalo, NY 14260, USA\\
\authormark{3}Institute for Quantum Computing and Department of Physics and Astronomy, University of Waterloo, Waterloo, ON N2L 3G1, Canada\\
\authormark{4}Division of Research and Innovation, University of Bristol, Bristol BS8 1QU, UK\\
\authormark{5}Azul Optics Ltd, Henleaze, Bristol BS9 4QG, UK}

\email{\authormark{*}g.misson@aston.ac.uk}

\begin{abstract*}
Optical anisotropy of the Henle fiber layer (HFL) underlies two macular
polarization phenomena, form birefringence (the macular cross) and dichroism
(Haidinger's brushes), which have conventionally been treated as unrelated. We
model the HFL as a radial diattenuating retarder and derive a single complex
dielectric tensor: effective-medium theory for parallel cylinders gives the
real part, and cylindrical averaging of the measured transmembrane xanthophyll
tilt gives the imaginary part. The birefringent slow axis is radial and the
absorption axis tangential, perpendicular by structural necessity. The model
reproduces measured dichroic ratios of 1.04--1.14 for a single calibrated
oriented-pigment fraction, and supplies the input parameters for full
electromagnetic treatments.
\end{abstract*}



\section{Introduction}
\label{sec:intro}

The interaction of polarized light with the human eye spans two
complementary domains: an objective domain, in which
polarization-dependent tissue properties are measured instrumentally, and a
perceptual domain, in which polarization-modulated stimuli give rise
to subjective entoptic phenomena \cite{temple2024}.  Two anatomical
structures dominate these interactions: the cornea, a birefringent anterior
element whose retardance modifies polarization states entering and leaving
the eye \cite{vanblokland1987,misson2010,knighton2002,shute1974}, and the
retina, within which the retinal nerve fiber layer and the Henle fiber layer
(HFL) of the central macula are the principal sources of measurable
polarization effects \cite{elsner2008,brink1988}.  The HFL is the focus of
the present work.

The macula exhibits two distinctive polarization-dependent phenomena:
Haidinger's brushes and the macular cross seen in retinal polarimetry.
These have traditionally been interpreted separately, Haidinger's brushes as
selective absorption by macular pigment and the macular cross as
birefringence of the HFL.  Yet both originate in the same radial and
tangential directions imposed by the radially arranged cone axons and
M\"{u}ller-cell processes of the HFL.  This shared geometry motivates the
central tenet of this work: that the HFL is a single radial
diattenuating retarder, a radially symmetric anisotropic element that
simultaneously imposes phase retardation (birefringence) and
polarization-dependent absorption (dichroism) along common principal axes.
Its natural constitutive description is therefore a single complex
dielectric tensor whose real part encodes the retardance and whose imaginary
part encodes the diattenuation, the macular cross and Haidinger's brushes
being complementary manifestations of this one element that reveal its real
and imaginary parts respectively.

The fovea centralis is a pit-like depression in the central macula that
maximizes spatial acuity by combining a high cone density with lateral
displacement of the inner retinal layers, yielding a short, low-scatter
optical path along the visual axis \cite{bringmann2018primate}.  As a
direct consequence of foveal pit formation, cone axons and their
accompanying M\"{u}ller cells run radially outward from the centrally
located photoreceptors to the peripherally shifted outer plexiform layer,
forming the HFL (Fig.~\ref{fig:anatomy}a--c).  These fibers are locally
parallel cylindrical units whose cytoplasm is optically distinct from the
surrounding extracellular matrix.  Molecules of the xanthophyll carotenoids lutein,
zeaxanthin, and meso-zeaxanthin span the full thickness of the cell membrane
\cite{bone1992a,bernstein2016} (although lutein may also adopt an
orientation parallel to the membrane; \cite{sujak1999,widomska2023}),
accumulating preferentially in the HFL
\cite{bone1984,snodderly1984a,curcio2018} and imparting the yellowish
color of the \emph{macula lutea}.  This radial organization produces an
azimuthally uniform optical anisotropy that is the structural basis of all
the phenomena considered here.

The radial anisotropy of the HFL has two structurally related optical
consequences.  First, the micron-scale cylindrical geometry generates
form birefringence: the refractive index along the fiber axis
($n_\parallel$) exceeds that transverse to it ($n_\perp$), so the HFL acts
as a radially oriented uniaxial retarder with its slow axis along the fiber
\cite{hemenger1982,brink1988}.  Second, the transmembrane orientation of
the xanthophyll chromophores generates macular dichroism: averaged
over the cylindrical membrane surface, the effective absorption dipole
points perpendicular to the fiber axis, preferentially absorbing
light polarized tangentially to foveal concentric circles.  The two
anisotropies thus share a common radial origin but act in orthogonal senses,
the birefringent slow axis lying along the fiber and the dichroic absorption
axis across it.

These anisotropies underlie the eye's best-known macular polarization
phenomena.  Haidinger's brushes \cite{haidinger1844}, the faint bow-tie
perceived in linearly polarized light, arise from macular dichroism 
; the macular cross, seen with crossed polarizers in
reflected-light fundus imaging \cite{hochheimer1978,hochheimer1982,brink1988}
and as the two-cycle ($2f$) retardance component in scanning laser
polarimetry \cite{elsner2008} and retinal birefringence scanning
\cite{hunter1999}, arises from form birefringence.  Because the birefringent
cornea (Fig.~\ref{fig:anatomy}a) is an anterior element optically separable
from the retina \cite{weinreb2003}, the intrinsic HFL dielectric tensor
derived here is obtained without it.

Despite an extensive experimental literature
\cite{bone1980,bone1984,brink1988,elsner2008}, birefringence and dichroism
have not previously been derived together from molecular structure, unified
within a single dielectric tensor, and expressed through their mutually
perpendicular principal axes in a physically interpretable ellipsoid.  That
tensor is the principal output of this work and provides the foundation for a full electromagnetic theory of macular optics. 

Following this introduction, Section~\ref{sec:geometry} establishes the foveal coordinate system and
models the HFL as a homogeneous anisotropic slab.  Section~\ref{sec:birefringence}
derives the real part of the tensor (form birefringence) from
effective-medium theory and calibrates the retardance against in-vivo measurements.
Section~\ref{sec:dichroism} derives the imaginary part (macular
dichroism) from the measured xanthophyll tilt angle \cite{grudzinski2017}
via cylindrical averaging, orientational disorder, and a sensitivity
analysis of the dichroic ratio.  Section~\ref{sec:tensor} assembles the full
complex dielectric tensor and introduces the dielectric ellipsoid, and
Sections~\ref{sec:discussion}--\ref{sec:conclusions} discuss and summarize
the results.

\paragraph{Scope.}
The analysis is restricted to monochromatic light at $\lambda=460\,$nm (the
macular-pigment absorption peak) and normal incidence, except where
$\lambda=514\,$nm is used for comparison with the Brink \& van Blokland
measurements.  


\section{Ocular Anatomy, Coordinate System and Slab Model}
\label{sec:geometry}

\begin{figure}[htbp]
\centering
\includegraphics[width=0.77\textwidth]{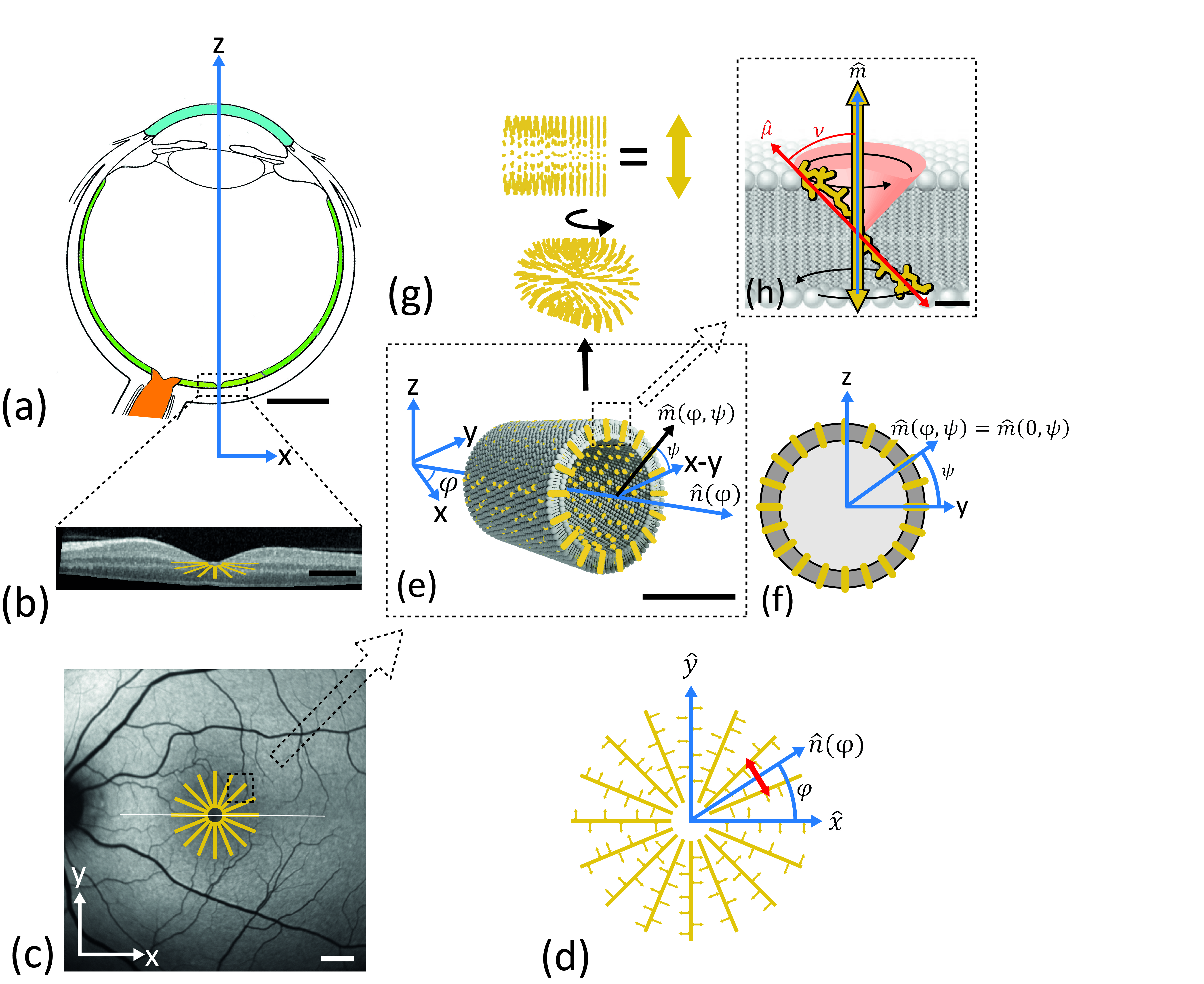}
\caption{\footnotesize%
  \textbf{Anatomy and geometry of the Henle fiber layer.}
  \textbf{(a)}~Schematic cross-section of the human eye (left eye viewed
  from below along the $y$-axis).  The visual ($-z$) axis runs from the
  cornea (cyan) through the lens and vitreous to the fovea.  The retina
  (green) lines the internal part of the eye behind the lens; the macula
  (light green) is the specialized central region of the retina.  The
  $x$--$y$ plane lies on the retinal surface centered at the fovea, with
  the $x$-axis horizontal between the fovea and the nearest edge of the
  optic nerve (orange).  [scale bar $=5\,$mm]
  \textbf{(b)}~Cross-section along $x$ through the macula revealed by
  optical coherence tomography (OCT), demonstrating the layered retinal
  architecture and the central foveal pit.  Yellow lines indicate the
  approximate location and lateral extent of the Henle fiber layer.  The
  layer thickness $h$ at any eccentricity is the parameter entering
  Eq.~(\ref{eq:retardance}).  [scale bar $=1\,$mm]
  \textbf{(c)}~Image of the retina as seen by ophthalmoscopy along the
  visual ($-z$) axis.  The OCT cross-section (b) is taken along the white
  line.  The yellow spokes represent the location of the radially
  arranged Henle fibers, each spoke corresponding to the local fiber
  direction $\nhat(\vp)=(\cos\vp,\sin\vp,0)$.  Retinal blood vessels
  (dark tortuous linear structures) converge on the optic disc midway
  along the left boundary.  [scale bar $=1\,$mm]
  \textbf{(d)}~Enlarged Henle fiber schematic showing the foveal
  coordinate geometry (view along $-\hat{z}$, as seen by the observer
  looking into the eye).  The yellow lines represent Henle fibers at
  regularly spaced azimuths.  At a given azimuth $\vp$, the unit fiber
  direction vector $\nhat(\vp)=(\cos\vp,\sin\vp,0)$ (blue arrow) runs
  radially from the foveal center and the effective absorption axis
  $\dhat(\vp)=(-\sin\vp,\cos\vp,0)$ (red arrow) is perpendicular to it
  and tangential to foveal concentric circles; the derivation of $\dhat$
  is given in Section~\ref{sec:dichroism}.  The angle $\vp$ between
  $\nhat(\vp)$ and $+\hat{x}$ is marked.  Together, $\nhat(\vp)$ and
  $\dhat(\vp)$ define the two principal optical axes of the HFL at every
  foveal location.
  \textbf{(e)}~Nano-scale diagrammatic representation of a Henle fiber
  (gray tube) with long axis $\nhat$ aligned at angle $\vp$.  The macular
  pigment molecules span the cell membrane, and their average alignments
  (yellow ellipses) follow the Henle fiber unit normal vectors $\hat{m}$,
  making an angle $\psi$ from the $x$--$y$ plane.  [scale bar $=1\,\mu$m]
  \textbf{(f)}~As (e) viewed along the $x$-axis ($\vp=0$).
  \textbf{(g)}~Diagrammatic demonstration of the bulk effect of the
  cylindrical arrangement of many molecular dipoles oriented along the
  Henle fiber surface normals: the origin of the effective absorption
  axis, drawn as tangential yellow arrows perpendicular to the Henle fiber
  (the same absorption direction $\dhat$ marked in panel~(d); see
  Section~\ref{sec:dichroism}).
  \textbf{(h)}~Molecular-scale representation of a macular pigment
  molecule spanning the Henle fiber cell membrane (after
  Grudzinski \textit{et al.} \cite{grudzinski2017}).  The molecular axis $\hat{\mu}$ makes an
  angle $\nu$ with the membrane normal $\hat{m}$.  Free rotation about
  $\hat{m}$ results in the average molecular dipole coinciding with
  $\hat{m}$.  [scale bar $=1\,$nm]
  The cylindrical-averaging construction in panels~(e--g) is partially
  adapted from Temple \textit{et al.} \cite{temple2019}.}
\label{fig:anatomy}
\end{figure}

Relevant ocular anatomy and a geometric summary are presented in Fig.~\ref{fig:anatomy}. A left eye is assumed, noting that a right eye is mirror-symmetric. Let the laboratory-frame origin be the foveal centre with $z$ aligned in the anterior direction along the visual axis (the opposite direction of light propagation), with
$x$ and $y$ spanning the retinal plane (Fig.~\ref{fig:anatomy}a,b,c).
The azimuthal angle $\vp$ is
measured counterclockwise from the positive $x$-axis (Fig.~\ref{fig:anatomy}d).  The Henle fiber
runs radially from the foveal center, so its unit direction vector at
azimuth $\vp$ is
\begin{equation}
  \nhat(\vp) = (\cos\vp,\,\sin\vp,\,0).
  \label{eq:fiber_dir}
\end{equation}
As derived in Section~\ref{sec:dichroism}, the effective optical
absorption axis, the net dipole direction after averaging over the
cylindrical membrane surface, is
\begin{equation}
  \dhat(\vp) = (-\sin\vp,\,\cos\vp,\,0),
  \label{eq:dipole_dir}
\end{equation}
perpendicular to $\nhat(\vp)$ and tangential to foveal concentric circles (Fig.~\ref{fig:anatomy}d).

The HFL is assumed to be a homogeneous anisotropic slab with thickness $h$
and constitutive relation $\mathbf{D} = \beps(\vp)\mathbf{E}$ , where $\mathbf{D}$, $\beps$, and $\mathbf{E}$ are the electric displacement field, the permittivity tensor, and the electric field, respectively.  Magnetic permeability is
taken as unity ($\mu_{ij} = \delta_{ij}$) and optical activity is
ignored.  At each azimuth $\vp$, the slab has one set of optical
principal axes defined by Eqs.~(\ref{eq:fiber_dir}) and
(\ref{eq:dipole_dir}).

The parameters and their numerical values used throughout
this paper are summarized in Table~\ref{tab:params}.  All symbols are defined at first use; a notation table and a
modeling-assumptions table are provided in Supplement 1
(Tables~S1 and~S2).

\paragraph{Notation conventions.}
Throughout this paper the following notation is used consistently for dichroic
ratio quantities: $\mathcal{R}=\mathrm{OD}_\perp/\mathrm{OD}_\parallel$ denotes
the measured (experimental) dichroic ratio; $R_{\rm geom}$ denotes the
geometric dichroic ratio arising from fiber geometry and orientational
averaging (a derived quantity); and $R_{\rm eff}$ denotes the effective
model dichroic ratio obtained after scaling by $f_{\rm oriented}$ (a fitted
quantity, constrained to reproduce $\mathcal{R}$).  The term
absorption axis is used throughout for the effective direction of
maximum optical absorption (synonymous with ``effective dipole direction'' or
``dichroic axis''), directed along $\dhat(\vp)$.

\begin{figure}[htbp]
\centering
\includegraphics[width=0.92\textwidth]{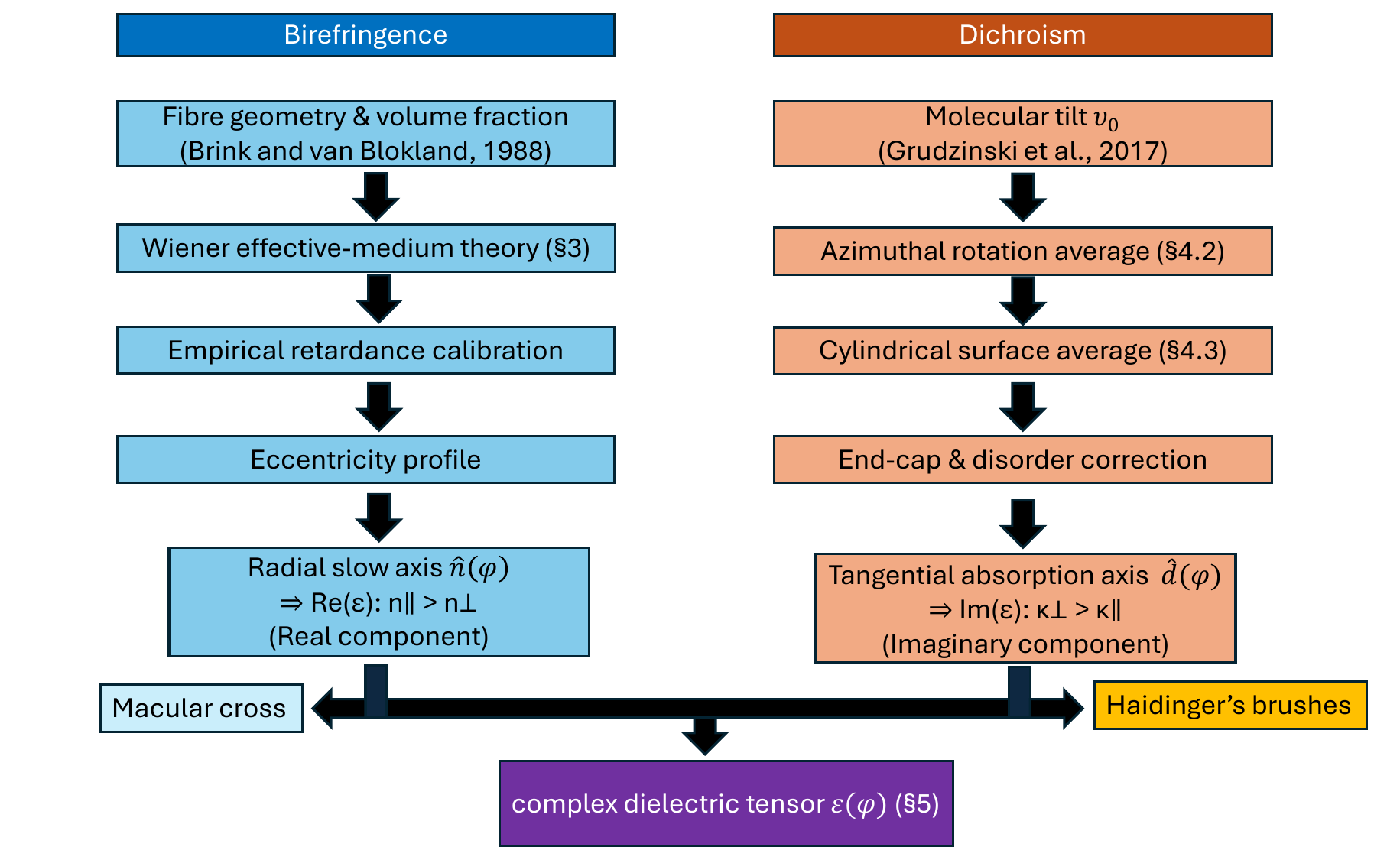}
\caption{%
  \textbf{Derivation roadmap: two parallel chains converging in the
  unified complex dielectric tensor.}
  \textit{Left chain (birefringence, real component, Section~\ref{sec:birefringence}):}
  Henle fiber geometry and volume fraction (from electron microscopy
  \cite{brink1988}) enter Wiener effective-medium theory
  (Section~\ref{sec:emt}) to establish the sign and order of magnitude of the
  form birefringence.  The numerical retardance is then calibrated
  against in-vivo polarimetric measurements and fitted to the Rayleigh
  eccentricity profile.  The output is the
  birefringent slow axis $\nhat(\vp)$, directed radially from the fovea,
  yielding $\Real(\beps)$: $n_\parallel > n_\perp$.  Applied through
  Jones calculus, this generates the
  macular cross.
  \textit{Right chain (dichroism, imaginary component, Section~\ref{sec:dichroism}):}
  Molecular tilt angle $\nu_0 = 42.3^\circ$ from
  Grudzinski \textit{et al.} \cite{grudzinski2017} enters a four-step averaging sequence
  (Section~\ref{sec:dichroism}): azimuthal rotation average, cylindrical
  lateral-surface average, end-cap and
  orientational-disorder correction (Supplement 1, Section~S3).  The
  output is the tangential absorption axis $\dhat(\vp)$, perpendicular
  to $\nhat(\vp)$ and tangential to foveal concentric circles, yielding
  $\Imag(\beps)$: $\kappa_\perp > \kappa_\parallel$.  This is the
  physical basis of Haidinger's brushes.
  %
  \textit{Convergence (Section~\ref{sec:tensor}):}
  Both chains enter the unified complex dielectric tensor
  $\beps(\vp) = \Real\bigl[\beps(\vp)\bigr] + i\,\Imag\bigl[\beps(\vp)\bigr]$,
  where $\Real(\beps)$ encodes birefringence and $\Imag(\beps)$ encodes
  dichroism (Eq.~\ref{eq:complex_eps}).
  }
\label{fig:roadmap}
\end{figure}

\paragraph{Derivation roadmap.}
The logical structure of the paper is summarized in
Figure~\ref{fig:roadmap}, which shows the two parallel derivation chains
that converge in the unified complex dielectric tensor.

\begin{table}[htbp]
\centering
\caption{Physical parameters for the Henle macular layer at
$\lambda=460\,$nm.  Birefringence parameters are derived in
Section~\ref{sec:birefringence}; molecular dichroism parameters in
Section~\ref{sec:dichroism}; macroscopic optical parameters in
Section~\ref{sec:tensor}.  Parameters are classified as:
\textbf{M}~=~measured (from cited experiment),
\textbf{D}~=~derived (calculated from M values),
\textbf{F}~=~fitted (phenomenological; constrained to reproduce experimental
range). \textsuperscript{*}$L$ denotes the near-horizontal central fiber
segment traversing the optical path, which sets the end-cap aspect ratio
$L/r$ in the dichroic-ratio derivation (Supplement 1, Section~S3); the full
histological Henle-fiber displacement \textsuperscript{**} is
$L_{\rm HF}=491$--$578\,\mu\mathrm{m}$.}
\label{tab:params}
\footnotesize
\setlength{\tabcolsep}{4pt}
\renewcommand{\arraystretch}{0.94}
\begin{tabularx}{\textwidth}{>{\raggedright\arraybackslash}p{4.0cm} >{$}l<{$} >{$}l<{$} >{\centering\arraybackslash}p{0.8cm} >{\raggedright\arraybackslash}X}
\toprule
Parameter & Symbol & Value & Type & Source \\
\midrule
\multicolumn{5}{l}{\textit{Molecular (Section~\ref{sec:dichroism})}} \\
Dipole tilt from membrane normal & \nu & 42.3^\circ & M & \cite{grudzinski2017} \\
Orientational order parameter & \delta_{\rm op} & 0.320 & D & Eq.~(\ref{eq:order_param}) \\
Dichroic anisotropy factor (ideal) & F & 1.71 & D & Supplement 1, Section~S3 \\
Central fiber segment length *, radius & L,r & 50,1\,\mu\mathrm{m} & D & assumed \\
Orientational disorder (S.D.) & \sigma_\nu & 12^\circ & F & assumed \\
Geometric dichroic ratio & R_{\rm geom} & 1.63 & D & Supplement 1, Section~S3 \\
Oriented fraction (simulation) & f_{\rm oriented} & 0.20 & F & Supplement 1, Section~S3 \\
\midrule
\multicolumn{5}{l}{\textit{Form birefringence (Section~\ref{sec:birefringence})}} \\
Fiber refractive index & n_f & 1.395 & M & \cite{brink1988} \\
Medium refractive index & n_m & 1.33 & M & \cite{brink1988} \\
Fiber volume fraction & f & 0.75 & M & \cite{brink1988} \\
Form birefringence & \Delta n_r & \approx 0.65\times 10^{-3} & F & \cite{brink1988} \\
Henle fiber length ** (histol.) & L_{\rm HF} & 491-578\,\mu\mathrm{m} & M & \cite{drasdo2007} \\
Henle layer thickness & h & 30\,\mu\mathrm{m} & M & \cite{brink1988} \\
Single-pass retardance (514\,nm) & \delta_r & \approx 13^\circ & D & Eq.~(\ref{eq:retardance}) \\
\midrule
\multicolumn{5}{l}{\textit{Macroscopic optical (Section~\ref{sec:tensor})}} \\
Background refractive index & n_b & 1.34 & M & \cite{bone1984} \\
Peak optical density (foveal) & D_0 & 0.399 & M & \cite{bone1984,misson2019} \\
Dichroic ratio (measured) & \mathcal{R} & 1.115 & M & \cite{devries1953,bone1980} \\
Max.\ transmittance ($T_\parallel$) & k_1 & 0.42 & M & \cite{bone1984} \\
Min.\ transmittance ($T_\perp$) & k_2 & 0.38 & M & \cite{bone1984} \\
Mean transmittance & A & 0.400 & D & \\
Dichroic coefficient & B & +0.020 & D & \\
Max.\ Haidinger contrast & |B|/A & 0.050 & D & Eq.~(\ref{eq:contrast_main}) \\
\bottomrule
\end{tabularx}
\end{table}




\section{Form Birefringence of the Henle Fiber Layer}
\label{sec:birefringence}
\label{sec:emt}

The HFL is both a circularly symmetric dichroic medium and also a circularly symmetric positively birefringent
medium. The two components arise from the same structural element consisting of the radially
oriented cone axons and Müller-cell processes whose cylindrical
cytoplasmic cores are embedded in lower-index extracellular material at
high volume fraction \cite{brink1988}.  Individual Henle fibers (Fig.~\ref{fig:anatomy}e; radius $r\approx1\,\mu$m,
Table~\ref{tab:params}) have cross-sections of the same order as the visible light waveband
or larger, so the layer lies outside the strict sub-wavelength regime in which
classical effective-medium theory is exact.  The Wiener expressions are
therefore used only to fix the sign and order of magnitude of the anisotropy,
with finite-size corrections (below) and empirical calibration supplying its
value.  In addition, the oriented membrane-bound pigment
molecules (Fig.~\ref{fig:anatomy}e--h) may contribute a small intrinsic birefringence via
Kramers--Kronig coupling between the dispersive and absorptive parts of
the refractive index \cite{brink1988}; this is implicitly included in
the complex dielectric tensor of Section~\ref{sec:tensor}.

We apply the classical Wiener effective-medium expressions for an array of
parallel cylindrical inclusions.  Let $n_f$, $n_m$, and $f$ denote,
respectively, the fiber refractive index, the surrounding medium index, and
the fiber volume fraction (numerical values in Table~\ref{tab:params}).
Effective-medium theory is fundamentally a theory for the permittivity; for
electric fields parallel and perpendicular to the cylinder axis the principal
permittivities are
\begin{align}
  \varepsilon_\parallel &\approx f\,\varepsilon_f + (1-f)\,\varepsilon_m,
  \label{eq:npar}\\
  \frac{1}{\varepsilon_\perp} &\approx \frac{f}{\varepsilon_f} + \frac{1-f}{\varepsilon_m},
  \label{eq:nperp}
\end{align}
with $\varepsilon_f=n_f^2$ and $\varepsilon_m=n_m^2$; the principal refractive
indices then follow as $n_{\parallel,\perp}=\sqrt{\varepsilon_{\parallel,\perp}}$.
Because the volume-weighted arithmetic mean (parallel) exceeds the harmonic
mean (perpendicular) whenever $\varepsilon_f>\varepsilon_m$, these give
$n_\parallel > n_\perp$ for all physically realistic values $n_f > n_m$,
implying positive birefringence with the slow axis aligned with the fiber
direction $\nhat(\vp)$.  The associated birefringence and
single-pass retardance are:
\begin{equation}
  \Delta n_r = n_\parallel - n_\perp,
  \label{eq:Dnr}
\end{equation}
\begin{equation}
  \delta_r(h) = \Delta n_r \cdot \frac{2\pi h}{\lambda}.
  \label{eq:retardance}
\end{equation}
Substituting the HFL parameters from electron microscopy
\cite{brink1988} ($n_f=1.395$, $n_m=1.33$, $f=0.75$) yields
$\Delta n_r \approx 1.2\times10^{-3}$, giving
$\delta_r \approx 25$--$28^\circ$ for $h = 30\,\mu$m at visible wavelengths.
As discussed below, finite-cylinder corrections reduce this raw
effective-medium estimate by roughly a factor of two, to the adopted value
$\Delta n_r \approx 0.65\times10^{-3}$ ($\delta_r \approx 13$--$15^\circ$).
The range in $\delta_r$ reflects the spread in wavelength ($460$--$514\,$nm)
across the in-vivo retardance data; the corresponding range in $\Delta n_r$
consistent with these data is approximately $(0.4$--$1.0)\times10^{-3}$,
bracketing the adopted value.

\paragraph{Limitation and empirical calibration.}
The Wiener expressions assume that refractive-index fluctuations are small
relative to the optical wavelength.  Individual Henle fibers, however, have
diameters comparable to or larger than
visible wavelengths, placing the HFL outside the strict asymptotic regime
of classical effective-medium theory.  Hemenger \cite{hemenger1989} showed, using a
second-order Born (Rayleigh--Gans) scattering approach, that finite-cylinder
corrections can reduce the predicted birefringence substantially, by
roughly half for retinal nerve fiber layer parameters, with the correction
growing as the ratio of fiber diameter to in-medium wavelength increases.

For this reason, effective-medium theory is used here only to establish the
sign and order of magnitude of the anisotropy, not its precise
numerical value.  The birefringence adopted in all subsequent calculations is
constrained by published in-vivo retardance measurements \cite{brink1988}
and should be regarded as an empirically calibrated effective parameter that
implicitly incorporates finite-size corrections, structural complexity, and
any intrinsic birefringence from oriented membrane constituents.  The value
$\Delta n_r \approx 0.65 \times 10^{-3}$ is consistent with the single-pass
retardations of $+13.5^\circ$ and $+12.5^\circ$ at $514$ and $568\,$nm that
Brink and van Blokland \cite{brink1988} estimate for the form birefringence of Henle's fiber layer,
and, after wavelength scaling, with the retinal birefringence-scanning data of
Hunter \textit{et al.} \cite{hunter1999} and the maximum double-pass HFL retardation of
$\approx22^\circ$ at $840\,$nm mapped by polarization-sensitive OCT, which
peaks near $1.8^\circ$ and is measurable only over $\approx1$--$5^\circ$ of
foveal eccentricity \cite{cense2013}; the last of these independently
confirms that the birefringence is confined to a limited eccentricity range.

The principal result of this section is therefore not a precise prediction
of $\Delta n_r$ but the identification of the fiber direction as the
birefringent slow axis (Eqs.~\ref{eq:npar}--\ref{eq:Dnr}), confirming its
mutual perpendicularity with the absorption axis derived subsequently in
Section~\ref{sec:dichroism}: a structural consequence examined further
in Section~\ref{sec:tensor}.


\section{Molecular Basis of Macular Dichroism}
\label{sec:dichroism}

The macular pigment consists of the xanthophyll carotenoids lutein,
zeaxanthin, and meso-zeaxanthin \cite{bone1992a,bernstein2016,wald1945}.  All three
are long-chain polyenes whose conjugated $\pi$-electron systems form the
light-absorbing chromophore.  Their preferential accumulation in the HFL
is well documented \cite{bone1984,snodderly1984a,bernstein2016,wald1945}.

Grudzinski \textit{et al.} \cite{grudzinski2017} measured the orientation of lutein and zeaxanthin
in single giant unilamellar vesicles formed from dimyristoylphosphatidylcholine
(DMPC) using three independent methods and concluded that both xanthophylls adopt a tilted
transmembrane orientation with the transition dipole at
$\nu\approx42^\circ$ from the membrane normal (Fig.~\ref{fig:anatomy}h).

Because the lipid bilayer is fluid, the molecule rotates freely about the
local membrane normal on timescales shorter than any optical measurement;
the optically relevant quantity is therefore the time-averaged
absorption tensor.

For a transition dipole $\hat{\mu}$ making angle $\nu$ with the membrane
normal $\hat{m}$ and rotating freely about $\hat{m}$, the time-averaged outer
product is:
\begin{equation}
  \langle\hat{\mu}_i\hat{\mu}_j\rangle
  = \cos^2\!\nu\;\hat{m}_i\hat{m}_j
    + \frac{\sin^2\!\nu}{2}\!\left(\delta_{ij} - \hat{m}_i\hat{m}_j\right).
  \label{eq:dyadic}
\end{equation}
Introducing the orientational order parameter
\begin{equation}
  \delta_{\rm op} \equiv \frac{3\cos^2\!\nu-1}{2}
            = \langle P_2(\cos\nu)\rangle,
  \label{eq:order_param}
\end{equation}
(the subscript ``op'' distinguishes this from the phase retardance $\delta_r$),
Eq.~(\ref{eq:dyadic}) becomes
\begin{equation}
  \langle\hat{\mu}_i\hat{\mu}_j\rangle
  = \frac{\sin^2\!\nu}{2}\,\delta_{ij}
    + \delta_{\rm op}\,\hat{m}_i\hat{m}_j.
  \label{eq:mu_avg}
\end{equation}
For the Grudzinski grand mean $\nu = 42.3^\circ$:
$\cos^2\!\nu\approx0.547$, giving $\delta_{\rm op}\approx0.320$.

We assume that only a fraction $f_{\rm oriented}$ of the macular-pigment
molecules is membrane-bound and therefore oriented, the remaining
$(1-f_{\rm oriented})$ being randomly oriented.  For the oriented population
the averaged tensor is Eq.~(\ref{eq:mu_avg}); for the random population the
average is isotropic, $\langle\hat{\mu}_i\hat{\mu}_j\rangle_{\rm random}=\tfrac13\delta_{ij}$.
The effective imaginary part of the dielectric tensor is the weighted sum of
these two contributions, with $\kappa_0$ the orientation-averaged (isotropic)
absorption strength of the chromophore:
\begin{equation}
  \begin{aligned}
    \varepsilon''_{ij}
    &= \kappa_0\!\left[
         \left(\frac{1-f_{\rm oriented}}{3}
           + f_{\rm oriented}\frac{\sin^2\!\nu}{2}\right)\!\delta_{ij}
         + f_{\rm oriented}\,\delta_{\rm op}\;\hat{m}_i\hat{m}_j
       \right] \\
    &\equiv \kappa_0\!\left[A_0\,\delta_{ij}
         + B_{\rm mol}\,\hat{m}_i\hat{m}_j\right],
  \end{aligned}
  \label{eq:eps_mol}
\end{equation}
Here,
\begin{align*}
  A_0 &= (1-f_{\rm oriented})/3 + f_{\rm oriented}\sin^2\!\nu/2,\\
  B_{\rm mol} &= f_{\rm oriented}\,\delta_{\rm op}.
\end{align*}

\paragraph{Physical interpretation of \texorpdfstring{$f_{\rm oriented}$}{f\_oriented}.}
Although $f_{\rm oriented}$ is calibrated phenomenologically to recover the
measured dichroic ratio, it has a direct physical interpretation: it is the
fraction of the total macular pigment that is preferentially (radially and
tangentially) organized and therefore contributes to polarization
filtering, the remaining $(1-f_{\rm oriented})$ being randomly oriented and
polarization-insensitive.  This is the same distinction drawn empirically
by Pushin \textit{et al.} \cite{pushin2025compod}, who separate the circularly oriented macular
pigment optical density (coMPOD) from the total MPOD using structured-light
entoptic stimuli; to first order $f_{\rm oriented}$ corresponds to the ratio
$\mathrm{coMPOD}/\mathrm{MPOD}$.  The two quantities are not identical, since
the dichroic strength $B_{\rm mol}=f_{\rm oriented}\,\delta_{\rm op}$ also
depends on the orientational order parameter $\delta_{\rm op}$: $f_{\rm
oriented}$ measures the fraction of pigment that is oriented, while
$\delta_{\rm op}$ measures the degree of order within that oriented
population.  The observation by Pushin \textit{et al.} \cite{pushin2025compod} that coMPOD
declines approximately as $1/r$ with eccentricity implies that a roughly constant
amount of oriented pigment is distributed around progressively larger
circumferences $2\pi r$. This provides an independent geometric basis for the
radial decline of the oriented-pigment contribution assumed in the spatial
profile of Section~\ref{sec:tensor}.

This geometric dilution need not be the sole cause.  The three macular
xanthophylls differ in both orientation and distribution: zeaxanthin and
meso-zeaxanthin adopt a predominantly transmembrane orientation, whereas
lutein can also lie parallel to the membrane plane
\cite{sujak1999,widomska2023}.  Because meso-zeaxanthin and zeaxanthin
predominate at the foveal center and lutein at greater eccentricities
\cite{bone1997}, the population-averaged orientational order, and hence the
effective dichroism, would be expected to fall with eccentricity
independently of fiber geometry.  This compositional effect is noted as a
caveat rather than incorporated into the single-tilt model, whose
transmembrane assumption is in any case not established in vivo, where the
xanthophylls may be held in a fixed orientation by a binding protein such as
StARD3 rather than partitioning freely into the bilayer.

Equation~(\ref{eq:eps_mol}) applies to a flat membrane with a fixed
normal $\hat{m}$.  The Henle axons are cylindrical: at azimuth $\vp$, the
local membrane normal at surface angle $\psi$ is
\begin{equation}
  \hat{m}(\vp,\psi) = -\sin\vp\cos\psi\,\hat{x}
                     + \cos\vp\cos\psi\,\hat{y}
                     + \sin\psi\,\hat{z}.
  \label{eq:mem_normal}
\end{equation}
Averaging Eq.~(\ref{eq:eps_mol}) over $\psi\in[0,2\pi)$ yields (see
Supplement 1, Section~S4):
\begin{equation}
  \langle\hat{m}_i\hat{m}_j\rangle_\psi
  = \frac{1}{2\pi}\!\int_0^{2\pi}
      \hat{m}_i(\vp,\psi)\,\hat{m}_j(\vp,\psi)\;d\psi
  = \frac{1}{2}\!\left(\delta_{ij} - \hat{n}_i\hat{n}_j\right).
  \label{eq:cylinder_avg}
\end{equation}
For light propagating along $\hat{z}$ (normal incidence), only the
transverse $xy$-block of the dielectric tensor couples $E_x$ to $E_y$.
The effective in-plane average reduces to a term proportional to
$\dhat\otimes\dhat$ (Supplement 1, Section~S4), giving the
effective transverse absorption tensor:
\begin{equation}
  \varepsilon''_\perp(\vp)
  = \kappa_0\!\left[A_0\,\mathbf{I}_2
    + \tfrac{1}{2}B_{\rm mol}\,\dhat(\vp)\otimes\dhat(\vp)\right],
  \qquad B_{\rm mol} = f_{\rm oriented}\,\delta_{\rm op}\geq0.
  \label{eq:eps_transverse}
\end{equation}
where $\mathbf{I}_2$ is the $2\times2$ identity matrix in the transverse
($xy$) plane.  This is the fiber-frame tensor $\diag(\varepsilon''_\parallel,\varepsilon''_\perp)$
with
\begin{equation}
  \varepsilon''_\perp = \kappa_0(A_0 + \tfrac{1}{2}B_{\rm mol})
  > \varepsilon''_\parallel = \kappa_0 A_0:
\end{equation}
absorption is greater perpendicular to the fiber (E-field parallel
to the dipoles), confirming the structural model of Bone \textit{et al.} \cite{bone1984}.

The geometry is illustrated in Fig.~\ref{fig:anatomy}(e--g).  The key result
of the cylindrical average is that the effective absorption axis is rotated:
it is no longer aligned with the local membrane normal but lies in the plane
perpendicular to the fiber axis.  This single geometric step converts the
molecular orientation into the macroscopic absorption axis of the Henle fiber
layer and underpins all subsequent optical calculations; the same geometric
translation was first pointed out by Temple \textit{et al.} \cite{temple2019}.

\paragraph{Dichroic ratio derivation.}
The four-step derivation of the effective dichroic ratio $R_{\rm eff}$
from the molecular tilt angle $\nu_0$ through cylindrical geometry
averaging, end-cap correction, orientational disorder, and oriented-fraction
scaling is given in full in Supplement 1 (Section~S4),
together with the sensitivity analysis (Supplement 1, Table~S4).
Setting $f_{\rm oriented}\approx0.06$--$0.22$ recovers the physiological
range $R_{\rm eff}=\mathcal{R}\approx1.04$--$1.14$
\cite{devries1953,bone1980}; the adopted foveal value $\mathcal{R}=1.115$
corresponds to $f_{\rm oriented}\approx0.20$.  Once the effective dichroic
ratio is known, the two principal transmittances of the layer follow as
$T_\parallel=A+B$ and $T_\perp=A-B$, where $A$ is the mean transmittance and
$B$ the dichroic coefficient (Table~\ref{tab:params}); the maximum Haidinger's
brush contrast is then
\begin{equation}
  \frac{|B|}{A} = \frac{T_\parallel-T_\perp}{T_\parallel+T_\perp} \approx 0.05.
  \label{eq:contrast_main}
\end{equation}

\section{Unified Complex Dielectric Tensor}
\label{sec:tensor}

The two derivations of Sections~\ref{sec:birefringence} and
\ref{sec:dichroism} can now be combined into a single complex
dielectric tensor.  In the fiber frame (fiber along $\hat{x}$, dipole along
$\hat{y}$), the full complex principal permittivities are:
\begin{equation}
  \varepsilon_{\parallel,\perp}
  = \varepsilon_b
    \pm \tfrac{1}{2}\Delta\varepsilon_r
    + i\,\kappa_{\parallel,\perp},
  \label{eq:complex_eps}
\end{equation}
where:
\begin{itemize}
  \item $\varepsilon_b = n_b^2$ with $n_b\approx1.34$ is the
    isotropic background permittivity; $n_b$ is the measured bulk foveal
    index \cite{bone1984} and sets only the mean level, the form
    birefringence entering through the increment
    $\pm\tfrac{1}{2}\Delta\varepsilon_r$ below, so the precise choice of
    $n_b$ (versus the effective-medium mean $\approx1.38$ of
    Eqs.~\ref{eq:npar}--\ref{eq:nperp}) does not affect the anisotropy;
  \item $\Delta\varepsilon_r = 2n_b\,\Delta n_r > 0$ encodes the
    form birefringence, giving
    $\Real(\varepsilon_\parallel) > \Real(\varepsilon_\perp)$
    (slow axis along the fiber, $n_\parallel > n_\perp$);
  \item $\kappa_{\parallel,\perp} = 2n_b k_{\parallel,\perp}$ are the
    absorptive imaginary parts, derived from the molecular anisotropy
    (Eq.~\ref{eq:eps_transverse}), with
    $\kappa_\perp > \kappa_\parallel$ (absorption axis perpendicular
    to the fiber, $k_\perp > k_\parallel$).
\end{itemize}
The full fiber-frame transverse tensor is therefore:
\begin{equation}
  \beps_{\rm fib}
  = \begin{pmatrix}\varepsilon_\parallel & 0 \\ 0 & \varepsilon_\perp\end{pmatrix}.
  \label{eq:eps_fib}
\end{equation}
The real (birefringent) and imaginary (dichroic) parts are here combined
phenomenologically; a fully causal, Kramers--Kronig-consistent coupling of
the dispersive and absorptive contributions is left to future work.

\begin{figure}[htbp]
\centering
\includegraphics[width=0.75\textwidth]{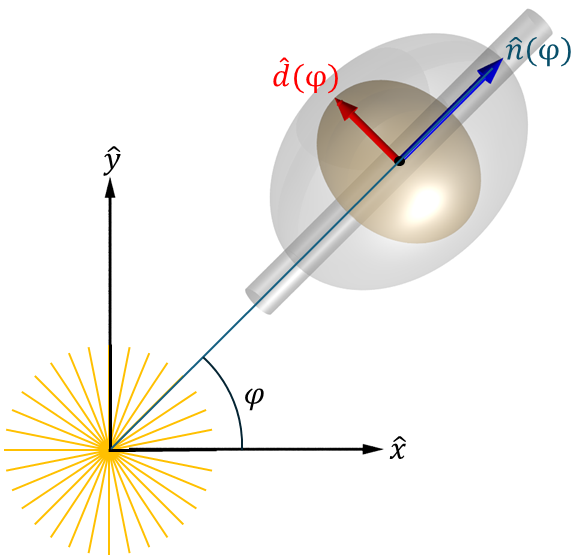}
\caption{%
  \textbf{The Henle fiber dielectric tensor ellipsoids projected onto the
  retinal plane} (view along $-\hat{z}$, visual axis into the page).
  An enlarged Henle fiber (gray cylinder) is oriented along $\nhat$ at an
  angle $\varphi$ from the unit vector $\hat{x}$.
  The gray ellipsoid (real part of $\beps$; birefringence) has its
  long (slow) axis along the fiber direction $\nhat$
  ($n_\parallel>n_\perp$; Eq.~\ref{eq:biref_sign}).
  The amber ellipsoid (imaginary part of $\beps$; dichroism) has
  its long axis along the dipole direction $\dhat$, perpendicular to the
  fiber ($\kappa_\perp>\kappa_\parallel$; Eq.~\ref{eq:dichroism_sign}).
  Dark blue arrow: fiber direction $\nhat(\varphi)$.
  Red arrow: effective absorption dipole $\dhat(\varphi)$.
  Collectively the Henle fibers are radially arrayed about the foveal
  center (yellow spokes).
  Ellipsoid aspect ratios are greatly exaggerated for visual clarity;
  the physical values are $\Delta n_r\approx0.65\times10^{-3}$ and
  $\mathcal{R}=\mathrm{OD}_\perp/\mathrm{OD}_\parallel\approx1.115$
  \cite{bone1984,devries1953}.  The $90^\circ$ locking of the
  birefringent slow axis (along $\nhat$) and the dichroic absorption
  axis (along $\dhat$) is a direct consequence of the cylindrical Henle
  fiber geometry (Sections~\ref{sec:dichroism} and~\ref{sec:emt}).}
\label{fig:ellipsoid}
\end{figure}

\paragraph{Perpendicular crossing of principal axes.}
Equation~(\ref{eq:complex_eps}) encodes a physically important asymmetry:
the real part $(\pm\tfrac{1}{2}\Delta\varepsilon_r)$ is positive along the
fiber ($\varepsilon_\parallel$) but negative perpendicular to it
($\varepsilon_\perp$), while the imaginary part
($i\kappa_{\parallel,\perp}$) is smaller along the fiber and larger
perpendicular to it.  Consequently:
\begin{equation}
  \Real(\varepsilon_\parallel) > \Real(\varepsilon_\perp)
  \quad\text{(slow axis}\parallel\text{fiber)},
  \label{eq:biref_sign}
\end{equation}
\begin{equation}
  \Imag(\varepsilon_\perp) > \Imag(\varepsilon_\parallel)
  \quad\text{(absorption axis}\perp\text{fiber)}.
  \label{eq:dichroism_sign}
\end{equation}
The radial fiber direction is thus simultaneously the birefringent slow axis
and the axis of maximum transmission, while maximum absorption lies along the
orthogonal tangential direction.  The $90^\circ$ locking of the two axes is a
derived consequence of this single radial diattenuating retarder, not a sign
of competing anisotropies, and is visualized in the dielectric ellipsoid
(Fig.~\ref{fig:ellipsoid}).

The imaginary parts are linked to measurable optical densities (OD):
\begin{equation}
  \mathrm{OD}_{\parallel,\perp} = \frac{\alpha_{\parallel,\perp}\,h}{\ln10},
  \qquad
  \alpha_{\parallel,\perp} = \frac{4\pi k_{\parallel,\perp}}{\lambda}.
  \label{eq:alpha}
\end{equation}
The dichroic ratio $\mathcal{R}=\mathrm{OD}_\perp/\mathrm{OD}_\parallel > 1$
gives:
\begin{equation}
  \mathrm{OD}_\perp = \frac{2\mathcal{R}\,D_0}{\mathcal{R}+1},\qquad
  \mathrm{OD}_\parallel = \frac{2D_0}{\mathcal{R}+1},
  \label{eq:OD}
\end{equation}
where $D_0 = (\mathrm{OD}_\perp + \mathrm{OD}_\parallel)/2$ is the peak
optical density.  The intensity transmittances $T_{\parallel,\perp}=10^{-\mathrm{OD}_{\parallel,\perp}}$
satisfy $T_\parallel > T_\perp$, with physiological values
$k_1=T_\parallel\approx0.42$ and $k_2=T_\perp\approx0.38$ at the foveal
center, derived from the in-vivo macular dichroic ratio and mean
transmittance reported by Bone \textit{et al.} \cite{bone1984} (Supplement 1, Section~S3).

The real parts are constrained by the form birefringence $\Delta n_r$
and, implicitly, by the Kramers--Kronig relations connecting the
wavelength-dependent dispersive and absorptive contributions from the
pigment chromophores.  Numerically, $\Delta\varepsilon_r = 2n_b\Delta n_r
\approx 2\times1.34\times0.65\times10^{-3} \approx 1.7\times10^{-3}$.

At foveal azimuth $\vp$, the fiber-frame tensor is rotated into the
laboratory frame by the standard orthogonal rotation $R(\vp)$:
$\beps_{\rm lab}(\vp) = R(\vp)\,\beps_{\rm fib}\,R(\vp)^\top$.  This
yields the $\vp$-dependent transverse ($xy$) block:
\begin{align}
  \varepsilon_{xx}(\vp) &= \varepsilon_{\rm avg}
    + \tfrac{1}{2}\Delta\varepsilon\cos2\vp, \label{eq:exx}\\
  \varepsilon_{yy}(\vp) &= \varepsilon_{\rm avg}
    - \tfrac{1}{2}\Delta\varepsilon\cos2\vp, \label{eq:eyy}\\
  \varepsilon_{xy}(\vp) &= \varepsilon_{yx}(\vp)
    = \tfrac{1}{2}\Delta\varepsilon\sin2\vp, \label{eq:exy}
\end{align}
where $\varepsilon_{\rm avg}=(\varepsilon_\parallel+\varepsilon_\perp)/2$
and $\Delta\varepsilon=\varepsilon_\parallel-\varepsilon_\perp$.  The
identity $\varepsilon_{xx}\varepsilon_{yy}-\varepsilon_{xy}^2
=\varepsilon_\parallel\varepsilon_\perp$ holds for all $\vp$, confirming
that the eigenvalues of the dielectric tensor are $\vp$-independent.

The off-diagonal term $\varepsilon_{xy}\propto\sin2\vp$ couples the
$x$- and $y$-polarization channels and is the direct geometric origin of
the sinusoidal modulation $I(\vp,\alpha)=A+B\cos2(\vp-\alpha)$
underlying Haidinger's brushes.  This term and its electromagnetic
consequences are developed elsewhere via the Berreman $4\times4$ matrix.

The real and imaginary parts of the complex dielectric tensor in the
fiber frame define two optical ellipsoids whose principal axes and aspect
ratios summarize the complete anisotropy of the Henle fiber layer
(Fig.~\ref{fig:ellipsoid}).  Projected onto the retinal plane as seen
along the visual axis ($-\hat{z}$, the direction of light propagation),
the gray ellipsoid (real part of $\beps$; birefringence) has its
long (slow) axis along the fiber direction $\nhat$
($n_\parallel>n_\perp$, Eq.~\ref{eq:biref_sign}), characterizing the
fiber as the optical slow axis, while the amber ellipsoid
(imaginary part of $\beps$; dichroism) has its long axis along the dipole
direction $\dhat$, perpendicular to the fiber
($\kappa_\perp>\kappa_\parallel$, Eq.~\ref{eq:dichroism_sign}).  The
projection makes the mutual perpendicularity of the two principal optical
axes directly apparent ($\nhat\perp\dhat$).

Because $R(\varphi)$ rotates both ellipsoids rigidly with the fiber at every
azimuth (Eqs.~\ref{eq:exx}--\ref{eq:exy}), the perpendicularity is preserved
at all foveal positions, the slow axis remaining radial and the absorption
axis tangential across the radially arrayed Henle fibers.

The physical values $\Delta n_r\approx0.65\times10^{-3}$ and
$\mathcal{R}\approx1.115$ establish that both ellipsoids are nearly
spherical in reality; the exaggerated aspect ratios in
Fig.~\ref{fig:ellipsoid} are chosen solely for visual clarity.

\section{Discussion}
\label{sec:discussion}

The principal contribution of this work is the demonstration that HFL
birefringence and macular dichroism are complementary manifestations of a
single anisotropic retinal structure, described mathematically by one
complex dielectric tensor.  The Henle fiber geometry generates a positive
form birefringence ($n_\parallel>n_\perp$, Eqs.~\ref{eq:npar}--\ref{eq:Dnr})
with the slow axis radial along $\nhat(\vp)$, the retardance increasing with
eccentricity \cite{elsner2008}; the same geometry, through cylindrical
averaging of membrane-bound xanthophyll orientation, produces a macroscopic
absorption axis tangential to foveal concentric circles
($\kappa_\perp>\kappa_\parallel$, Eq.~\ref{eq:eps_transverse}) whose
magnitude scales as $f_{\rm oriented}\delta_{\rm op}$.  The resulting
orthogonality of the retardance and absorption axes is therefore not
coincidental but a structural necessity encoded in the complex tensor
(Eq.~\ref{eq:complex_eps}) and visualized in the dielectric ellipsoid
(Fig.~\ref{fig:ellipsoid}), requiring no independent assumptions.

Within this representation birefringence and dichroism are the real and
imaginary parts of one constitutive description, linking molecular
organization to the observable polarization signatures: the macular cross
(from $\Real(\beps)$) and Haidinger's brushes (from $\Imag(\beps)$).  Because both observables originate in the same off-diagonal
coupling $\varepsilon_{xy}\propto\sin2\vp$ (Eq.~\ref{eq:exy}) yet probe
orthogonal tensor components, their patterns bear a fixed angular
relationship that constitutes a stringent consistency test of
the unified model, developed quantitatively elsewhere.

Several parameters are experimentally constrained rather than predicted
\emph{ab initio}: the effective birefringence $\Delta n_r$ is calibrated
against in-vivo retardance measurements, and $f_{\rm oriented}$ is
phenomenological.  The framework should therefore be read as a physically
motivated, experimentally constrained model rather than a fully predictive
microscopic theory.

\paragraph{Robustness of principal conclusions.}

The qualitative conclusions of this paper are geometric consequences of
HFL cylindrical symmetry and are independent of the specific parameter
values adopted. These conclusions are the mutual perpendicularity of the
birefringent slow axis and dichroic absorption axis, the intrinsic
fourfold symmetry of the macular cross (in the absence of corneal
birefringence), and the tangential orientation of the effective
absorption dipole. The sensitivity analysis (Supplement 1, Table~S4)
demonstrates that $R_{\rm eff}$ remains within the measured range $1.04$
to $1.14$ across the full physically plausible ranges of $\nu_0$,
$\sigma_\nu$, and $L/r$, with only $f_{\rm oriented}$ requiring
calibration against a single observable.  The birefringence-derived
retardance predictions are consistent with independent measurements at
two wavelengths and two experimental geometries.  The modeling assumptions and their expected impact are
summarized in Supplement 1, Table~S2.

The birefringent retardance of the Henle fiber layer
($\delta_r\approx13^\circ$ at $\lambda=514\,$nm for $h=30\,\mu$m) does not
affect the transmitted intensity at normal incidence in a parallel-polarizer
geometry: it modifies only the polarization state, not the total intensity
\cite{misson2003}.  Under crossed polarizers, however, the birefringence
produces the macular cross, whose intensity pattern is derived in full elsewhere.  The empirical observation that the slow
axis of the macular retarder is perpendicular to the absorption axis of the
macular polarizer \cite{brink1988} is precisely accounted for by
Eqs.~(\ref{eq:biref_sign}) and~(\ref{eq:dichroism_sign}).
The dichroic anisotropy is correspondingly a small modulation of the mean
transmittance ($A\approx0.40$); its perceptual consequences follow from
the Mueller-matrix treatment developed elsewhere.

Brink and van Blokland \cite{brink1988} noted that oriented lutein molecules within the Henle
fiber membranes could contribute an intrinsic birefringence in addition
to the structural form birefringence, since dichroism and birefringence
are Kramers--Kronig conjugates.  In the formalism of
Section~\ref{sec:tensor}, this intrinsic contribution is implicitly folded
into the complex dielectric tensor: the total real part
$\Real(\varepsilon_{\parallel,\perp})$ includes both the form birefringence
from Eqs.~(\ref{eq:npar})--(\ref{eq:Dnr}) and the dispersive contribution
from the chromophore absorption via the Kramers--Kronig integral.  At
$\lambda=460\,$nm, well inside the macular pigment absorption band, this
dispersive contribution could be of comparable magnitude to the form
birefringence, but its accurate quantification requires spectroscopic
modeling beyond the scope of the present work.

The effective-medium derivation assumes perfect cylindrical fibers with
uniform cross-section.  Near the foveal center, the pit geometry departs
substantially from this idealization, and fiber radius and length change
with eccentricity in a manner not captured by the fixed-thickness slab
model.  A more complete model would couple the effective-medium parameters
to eccentricity explicitly.  The smooth-medium treatment also omits scattering
from the wavelength-scale spacing and index contrast between individual
fibers, which the empirical calibration of $\Delta n_r$ absorbs only in an
averaged sense.  Additionally, the dichroic ratio derivation
is restricted to normal incidence ($\xi=0$); for oblique rays the effective
dipole direction shifts, and form dichroism from cylindrical symmetry
\cite{hemenger1982} may contribute to $\Delta\varepsilon_r$.
Extensions incorporating wavelength-dependent dispersion, Kramers--Kronig
spectroscopic analysis, oblique incidence, and spatially varying
microstructure are naturally accommodated within the tensor formalism
and deferred to future work.

Finally, the absolute retardance and dichroism scale with macular-pigment
density, which varies markedly between individuals, so the numerical values
adopted here are representative rather than universal.

\section{Conclusions}
\label{sec:conclusions}

A unified optical description of the Henle fiber layer has been developed,
connecting tissue microstructure, macular form birefringence, and retinal
birefringence within a common complex dielectric-tensor framework, and
deriving macular dichroism from molecular xanthophyll orientation.  The
analysis identifies the cylindrical Henle fiber geometry as the structural
origin of both the radial birefringent slow axis ($\Real(\beps)$) and, through
cylindrical averaging of membrane-bound chromophores, the tangential absorption
axis ($\Imag(\beps)$).  These two axes are mutually perpendicular by structural
necessity.  The resulting complex dielectric tensor provides a compact
description of HFL optical anisotropy that naturally generates the macular
cross and the polarization symmetries underlying Haidinger's
brushes, and provides the input parameters for the
full electromagnetic treatments to be developed elsewhere.

The principal conclusions are:

\begin{enumerate}

  \item Averaging the molecular absorption tensor of the xanthophyll
    chromophores over the cylindrical Henle fiber surface orients the
    effective dipole direction
    \[
      \dhat(\vp)=(-\sin\vp,\cos\vp,0),
    \]
    perpendicular to the fiber (Eq.~\ref{eq:eps_transverse}).
    This is the rigorous geometric origin of the Bone \textit{et al.} \cite{bone1984}
    structural model and gives $\kappa_\perp>\kappa_\parallel$
    (absorption axis $\perp$ fiber).

  \item A four-step molecular derivation converts the Grudzinski tilt angle
    $\nu=42.3^\circ$ into an effective dichroic ratio
    $R_{\rm eff}\approx1.04$--$1.14$ (Supplement 1, Section~S3),
    recovering the in-vitro measurements of de Vries \textit{et al.} \cite{devries1953} and
    Bone \textit{et al.} \cite{bone1980} with $f_{\rm oriented}\approx6$--$22\%$.  In the
    small-$f$ limit this reduces to $\mathcal{R}\approx1+\tfrac{3}{2}f\delta_{\rm op}$
    (Supplement 1, Section~S3).

  \item Effective medium theory for parallel cylinders
    (Eqs.~\ref{eq:npar}--\ref{eq:Dnr}) yields $n_\parallel>n_\perp$ (slow
    axis $\parallel$ fiber), with $\Delta n_r\approx0.65\times10^{-3}$
    from the fiber geometry of Brink and van Blokland \cite{brink1988}.  The predicted
    single-pass retardance of $\approx13^\circ$ matches the $+13.5^\circ$
    (single-pass, $514\,$nm) estimated from the in-vivo Mueller-matrix
    ellipsometry of Brink and van Blokland \cite{brink1988}, and the resulting birefringent
    pattern of the macular cross will be developed elsewhere for both
    single-pass and double-pass (fundus-photography) geometries.

  \item The slow axis of birefringence (parallel to fiber) and the
    absorption axis of dichroism (perpendicular to fiber) are mutually
    perpendicular.  This structural constraint is encoded in the complex
    dielectric tensor $\varepsilon_{\parallel,\perp}=\varepsilon_b
    \pm\tfrac{1}{2}\Delta\varepsilon_r+i\kappa_{\parallel,\perp}$
    and is visualized by the dielectric ellipsoid of
    Fig.~\ref{fig:ellipsoid}, which serves as the primary output
    of this paper.

  \item The lab-frame rotation of the fiber-frame tensor introduces the
    off-diagonal coupling $\varepsilon_{xy}\propto\sin2\vp$
    (Eq.~\ref{eq:exy}), which is the geometric origin of both the
    macular cross (birefringence) and Haidinger's brushes (dichroism).

\end{enumerate}


\begin{backmatter}
\bmsection{Funding}
D.A.P is supported by the Natural Sciences and Engineering Research Council of Canada grant [RGPIN-2024-05220] and the Canada First Research Excellence Fund


\bmsection{Disclosures}
D.A.P. and D.S. are founders of Incoherent Vision Inc. S.E.T. is founder of Azul Optics Ltd. Both companies develop devices for polarization perception. The three authors listed as founders of the two companies are the major shareholders of their companies. D.A.P. and D.S. have patents on using structured light for creating entoptic profiles. S.E.T. has patents on using polarized light for creating entoptic profiles.


\bmsection{Data availability}
No data were generated or analyzed in the presented research.

\bmsection{Supplemental document}
See Supplement 1 for supporting content.

\end{backmatter}


\bibliography{haidinger_v4}

\end{document}